\documentclass[12pt]{article}
\usepackage{graphicx}
\begin{document}
\centerline{\bf Urban segregation with cheap and expensive residences}

\bigskip
\centerline{M.A. Sumour$^1$, A.H. El-Astal$^1$, M.A. Radwan$^1$, M.M. Shabat$^2$.} 
\bigskip

\noindent
$^1$ Physics Department, Al-Aqsa University, P.O.4051, Gaza, Gaza Strip, 
Palestinian Authority.

\noindent
$^2$ Physics Department, Islamic University, P.O.108, Gaza, Gaza Strip,
Palestinian Authority. Presently at Max Planck Institute for the Physics of
Complex Systems, N\"othnitzer Strasse 38, 01187 Dresden, Germany
\medskip

e-mail: msumoor@alaqsa.edu.ps

\bigskip

{\bf Abstract:}
In this paper we study urban segregation of two different communities A and B, poor and rich, distributed randomly on finite samples, to check cheap and expensive residences.  For this purpose we avoid the complications of the Schelling model which are not necessary and instead we use the Ising model on $500 \times 500$ square lattice, which give similar results, with random magnetic field at lower and higher temperatures ( $k_B T / J = 2.0, 99.0$) in finite times equal to 40, 400, 4000 and 40,000.  This random-field Ising magnet is a suitable model, where each site of the square lattice carries a magnetic field $\pm h$ which is randomly up (expensive) or down (cheap).  The resulting addition to the energy prefers up spins on the expensive and down spins on the cheap sites.  Our simulations were carried out using a 50-lines FORTRAN program.  We present at a lower temperature (2.0) a time series of pictures, separating growing from non-growing domains.  A small random field ($h = \pm 0.1$) allows for large domains, while a large random field ($h = \pm 0.9$) allows only small clusters.  At higher temperature (99.0) we could not obtain growing domains.


\bigskip
{\bf \large Introduction}

Sociophysics is the study of social questions by physicists using their physics methods.  In contrast to biophysics, it is a field which is not yet very well established. Opinion dynamics is one of the most widespread topics of sociophysics.
Statistical physics plays a useful rule, and social scientists \cite{schelling} have applied it, without knowing then that they dealt with an Ising model of ferromagnets. It is not at all the merit (or ignorance) of physicists which treats humans like numbers; this method has a very long tradition and is an indispensable part of modern life. 

The Schelling model of 1971 is a complicated version of a square-lattice Ising model at zero temperature, to explain urban segregation with two groups A and B, based on the neighbour preference of the residents, without external reasons.
Schelling published in 1971\cite{schelling}, in the same year in which physicist Weidlich  \cite{weidlich} published his first paper on sociodynamics.  Schelling in 2005 got the economics "Nobel" prize; his 1971 paper has an exponentially rising citation rate and is second-most-cited paper in this Journal of Mathematical Sociology.
In the original Schelling model, unhappy people move to the closest empty residences where they are happy, with 'happy' being defined as having at least half of the neighbours coming from the own group, and 'unhappy' meaning that the majority of neighbours comes from the other group.  This version leads to small clusters \cite{schulze,kirman,solomon} but not to large domains as for blacks and whites in Harlem, New York City \cite{sethi}.  Removing and replacing some people randomly \cite{jones} helps, and one also gets 'infinitely' large domains if happy (unhappy) people move to other places where they are happy (unhappy) \cite{kirman}.
Schelling\cite{schelling} asked whether the racial segregation in American cities can
emerge from intrinsic behaviour of the individual people, instead of or in
addition to extrinsic reasons like discrimination, rent differences, etc. In particular, can 'black' ghettos in the predominantly 'white' USA arise just because people prefer to have neighbours of their own group over neighbours from the other group?  In many other countries we find many other types of residential segregation, based on religion, ethnicity,.... In physics, such a process is easily simulated through the two-dimensional Ising model.

More plausible is a finite temperature $T$ \cite{ortmanns,schulze,solomon}, where people sometimes  move from happy to unhappy places, e.g. because they got a new job elsewhere. 
A new feedback between the composition of the neighbourhood and a local $T$ gives a self-organised $\langle T \rangle$ which may be on one or the other side of a phase transition between the two groups A and B \cite{muller}. This work 
was continued by Dall'Asta et al., using the Vincovic-Kirman version \cite{asta}, and by Odor \cite{odor} using one $T_1$ for tolerance and another $T_2$ for external noise.
The social meaning of temperature $T$ is not what we hear in the weather reports but an overall approximation for all the more or less random events which influence our decisions.  For residential segregation the model only counts how many neighbours of which group one has. But not all people of one group are alike, housing in different parts of a city costs different amounts of money, some parts are more beautiful than others, and job hunting may force us into a temporary residence of a new city which does not conform to our wishes.  In this way, a positive temperature allows for rare moves which increase the energy, i.e. we move to a new residence where the neighbourhood composition alone makes us less happy.  At zero temperature, the Ising model does not properly order into one or two â'infinite' domains.

Thus the temperature here can have two meanings: tolerance and noise. Tolerance means that for high $T$ one is willing to live among neighbours from a different group, and low $T$ means that he strongly prefers to live among neighbours of his own group.  The alternative interpretation is $T$ = noise; $T$ then measures all those facts of life outside the model which force people to move to another residence even though they like their old residence better.
While then the domains in the Schelling model at positive temperature seem to grow towards infinity \cite{solomon}, it is simpler to achieve the same aim in the well-known two-dimensional Ising model, with or without conserved number of people within each group. 

Some researchers\cite{schulze,ortmanns} implemented a suggestion of Weidlich \cite{weidlich} that people slowly learn to live together with neighbours from the other group.  Thus $T$ not only takes into account the various accidents from outside the model, but also measures the tolerance: The higher $T$ is the more are people willing to live in neighbourhoods of the other group.  In the limit $T$ = infinity the neighbours would not matter at all, for intermediate $T$, we get small clusters but no large domains, and for low $T$ the domains grow to infinite sizes on an infinite lattice.  The learning suggested by Weidlich thus means that this parameter $T$ (= temperature or tolerance) no longer is kept constant but slowly increases.
For an Ising model, \cite{ortmanns} showed how an initial large domain dissolves if the temperature is slowly increased from below to above $T_c$.  More realistically, for five (instead of only two) different groups in a modified 5-state Potts model, \cite{schulze} increased $T$ from low to high values and showed that with a slow increase one has appreciable domain formation during intermediate times, while with a fast increase this segregation is mostly avoided.
Unfortunately, sociologists ignored for decades the Ising model (and also barely cited \cite{jones}) while physicists until 2000 mostly ignored the Schelling model.

In the Schelling-Ising model\cite{schelling} of urban segregation, all lattice sites are equivalent. In reality, some houses are cheap and others are expensive\cite{benard}. And usually of two groups in a population, one is poorer than the other.  These models use two groups A and B of people (or more than two \cite{schulze}, distributed on a square lattice, with group A goes to up spins ($+1$), while group B goes to down spin ($-1$), and assume that everybody prefers to be surrounded by lattice neighbours from the same group instead of from the other group. Flipping a spin then means that a person from one group leaves town and is replaced by a person from the other group, as simulated already by\cite{jones}. 

Often, a population can be approximately divided into rich and poor \cite{benard}. Starving associate professors and luxuriously living full professors are one example, but poorer immigrants and richer natives are more widespread.  Poor people cannot afford to live in expensive houses, but rich people can.  If there are whole neighbourhoods of expensive and cheap housing, then these housing conditions enforce a segregation of rich from poor, and this segregation does not emerge in a self-organized way. The more interesting case allows for self-organization of domains by assuming that each residence randomly is either expensive or cheap, with no spatial correlations in the prices.

In this paper we use the random-field Ising model to distinguish between cheap and expensive residences at one $T$. We present simulation at lower and higher 
temperatures for different finite times.  
For such purpose we give a complete 50-line FORTRAN program.
Parts of the text and program were taken from \cite{stauffer,muller}

\bigskip
\centerline{\bf \large Random-Field Ising Model}

In the Ising model, two neighbouring spins have due to their interaction $-J S_i S_k$ 
a higher probability to belong to the same group than to belong to
the two different groups.  If the difference between these two probabilities is large enough,
$T<T_c$, domain sizes can grow to infinity in an infinite lattice.  While only small clusters are formed for smaller differences in the probabilities, $T>T_c$.
That these probabilities, controlled through $-J/k_B T$, lead to these different regimes, separated by a sharp phase transition at $T = T_c$, is not obvious from the definition of the interaction  $-J S_i S_k$, took physicists many years to find, and is typical of complex systems.
The earlier standard Ising model gives results similar to the properly modified
Schelling model \cite{jones,kirman,solomon}.

The magnetic field is considered to refer to the price of residence: people living in expensive residence are presented by positive magnetic field $h > 0$, or in cheap residence presented by negative magnetic field $ h < 0$.  Group A prefers to go to the cheap residences and group B to the expensive residences.  This can be simulated by a random magnetic field which is $+h$ on half of the places (attracting people of group A = up spins) and is $-h$ on the other half of the places (attracting group B = down spins).  The signs of the field are distributed randomly, and the model is called the random-field Ising model. This random field Ising model has a long history, and the asymptotic behavior for infinite system and infinite times is known. But for sociophysics we need finite systems and finite times and we use $500 \times 500$ lattices and hundreds of time steps only. We simulate to what extent the two groups segregate.

In the spin 1/2 Ising model with Glauber kinetics of the square lattice, we interpret the two spin orientations as representing two groups of people.  Nearest neighbours are coupled ferromagnetically, i.e. people prefer to be surrounded by others of the same group and not of the other group.  Starting with random initial distribution of zero magnetization (= number of one group minus number of the other group), we check if "infinitely" large domain are formed.  It is well known that they do so in zero magnetic field for $0 < T < T_c$ where $T_c = 2.269$ is the critical temperature in units of the interaction energy.
In this Ising model at finite $T$, each pair $< i, k >$ of nearest neighbours produces an energy $-J S_i S_k$ with some proportionality constant $J$. 
The total energy $E$ (= total unhappiness) is the sum of this pair energy over all neighbour pairs of the lattice. In statistical physics, different distributions of the spins $S_i$ are
realized with a probability proportional to exp$(-E/k_B T)$ where $T$ is the absolute temperature and $k_B$ the Boltzmann constant.
There is no need to worry about values for $T, k_B, J$ since the only relevant quantity is the ratio $k_BT/J$, taken as $2$ and $99.0$ in our simulations. The Glauber kinetics is simulated on the computer by flipping a spin if and only if a random number between
0 and 1 is smaller than the probability exp$ (-\Delta E/k_BT)/[1 +\exp(-\Delta E/k_B T)]$, where $\Delta E$ is the energy change produced by this spin flip. 
In addition we use a random field $\pm h$; the probabilities proportional to exp (--Energy$/k_BT)$ now depend also on the (local) field by a factor exp(Field/$k_BT$).
The Fortran program used in this study contains 50 lines, takes a few seconds, and is presented below:
\begin{verbatim}
      parameter(L=500,Lmax=(L+2)*L,L2=L*L)
      dimension is(Lmax),iex(9),iex1(9),iex2(9),h(Lmax)
      byte is
      data T,max,h0,ibm/2.0,400000,0.90,1/
      print *,'# L,max,ibm,t,h0',L,max,ibm,t,h0
      Lp1=L+1
      L2pL=L2+L
      do 1 i=1,Lmax
      is(i)=-1
      ibm=ibm*16807
1      if(ibm.gt.0) is(i)=1
      do 10 i=1,Lmax
      h(i)=-h0
      ibm=ibm*16807
10      if(ibm.gt.0) h(i)=h0
         do 2 ie=1,9
        ex1=exp(-2*((ie-5+h0))/T)
2        iex1(ie)=(2.0*ex1/(1.0+ex1) - 1.0)*2147483647
        do 20 ie=1,9
          ex2=exp(-2*((ie-5-h0))/T)
20        iex2(ie)=(2.0*ex2/(1.0+ex2) - 1. 0)*2147483647
c         print *, '#iex1', iex1,iex2
          ibm=2*ibm+1
          do 3 mc=1,max
          do 4 i=Lp1,L2pL
          ie=5+IS(I)*(is(i-1)+is(i+1)+is(i-L)+is(i+L))
          ibm=ibm*16807
           if(h(i)*is(i).lt.0) then
           if(ibm.lt.iex2(ie)) is(i)=-is(i)
            else
           if(ibm.lt.iex1(ie)) is(i)=-is(i)
           endif
4         continue
           mag1=0
           mag2=0
      do 6 i=Lp1,L2pL
       if(h(i).gt.0) mag1=mag1+is(I)
      if(h(i).lt.0) mag2=mag2+is(I)
c3     if(mc.eq.(mc/100000)*100000) print *, mc,mag
6      mag=mag1+mag2
c3       print *, mc,mag,mag1,mag2
3     continue
      do 5 i=Lp1,L2pL
      if(is(i).ne.1) goto 5
      iy = (i-1)/L
      ix=i-L*iy
       print *, ix, iy
5     continue
      stop
      end
\end{verbatim}

\bigskip

Such models and programs are taught in courses on computational or theoretical physics all over the world; the model was published in 1925. If in the above flipping probability the denominator is omitted one gets the Metropolis kinetics.  If instead of flipping one spin we exchange two opposite spins, we get the Kawasaki dynamics.  For Glauber or Metropolis, after very long times (measured by the number of sweeps through the lattice) one of the two possibilities dominates at the end, if $T$ is not larger that the critical temperature $T_c$, with $2J/ k_B T_c = \ln(1 + \sqrt 2) \simeq 0.88 $ known since 1940.

Various versions between Ising and Schelling models give about the same results
\cite{muller}. We therefore use in the present work the simple standard two-dimensional Ising model with Glauber dynamics instead of the complicated Schelling model.  In this study, initially we carry out our simulation at higher temperature of $T = 99.0$, with size of square lattice of $500 \times 500$, at time = 400, and vary the value of the field from 1.0 to 300.0 to check that our program agrees with the exact result for non-interacting spins.  No large domains were noticed at high temperature of 99.0.  It was also demonstrated that there are no growing domains obtained at high field.
We then tested the growing of domains of groups in small random field of $\pm 0.1$, and low temperature of 2.0 for at different times (40, 400, 4000, 40000) and the results of this simulation are presented in Figure 1(a,b,c,d).

\begin{figure}[hbt]
\begin{center}
\includegraphics [angle=-90,scale=0.3]{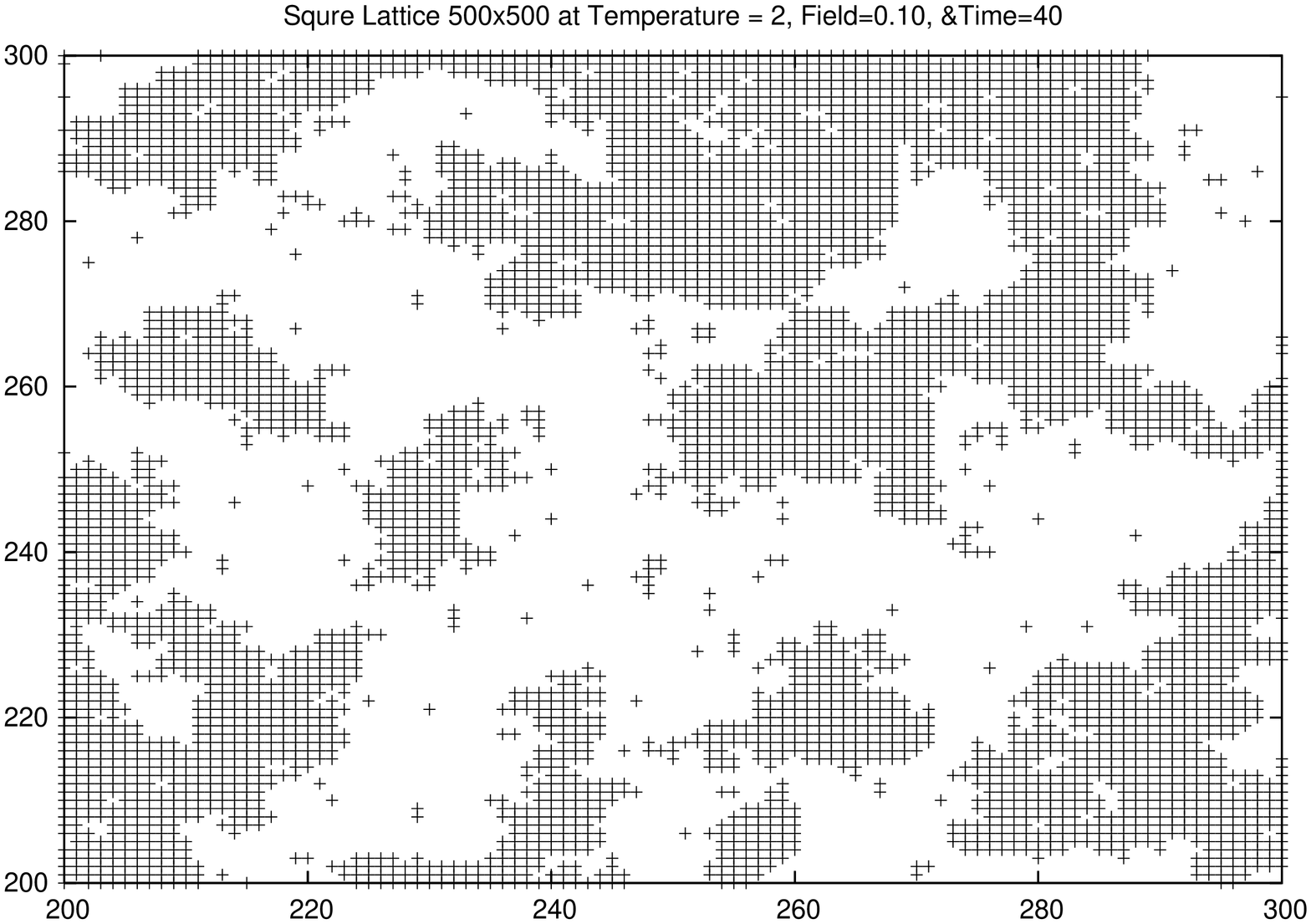}
\includegraphics [angle=-90,scale=0.3]{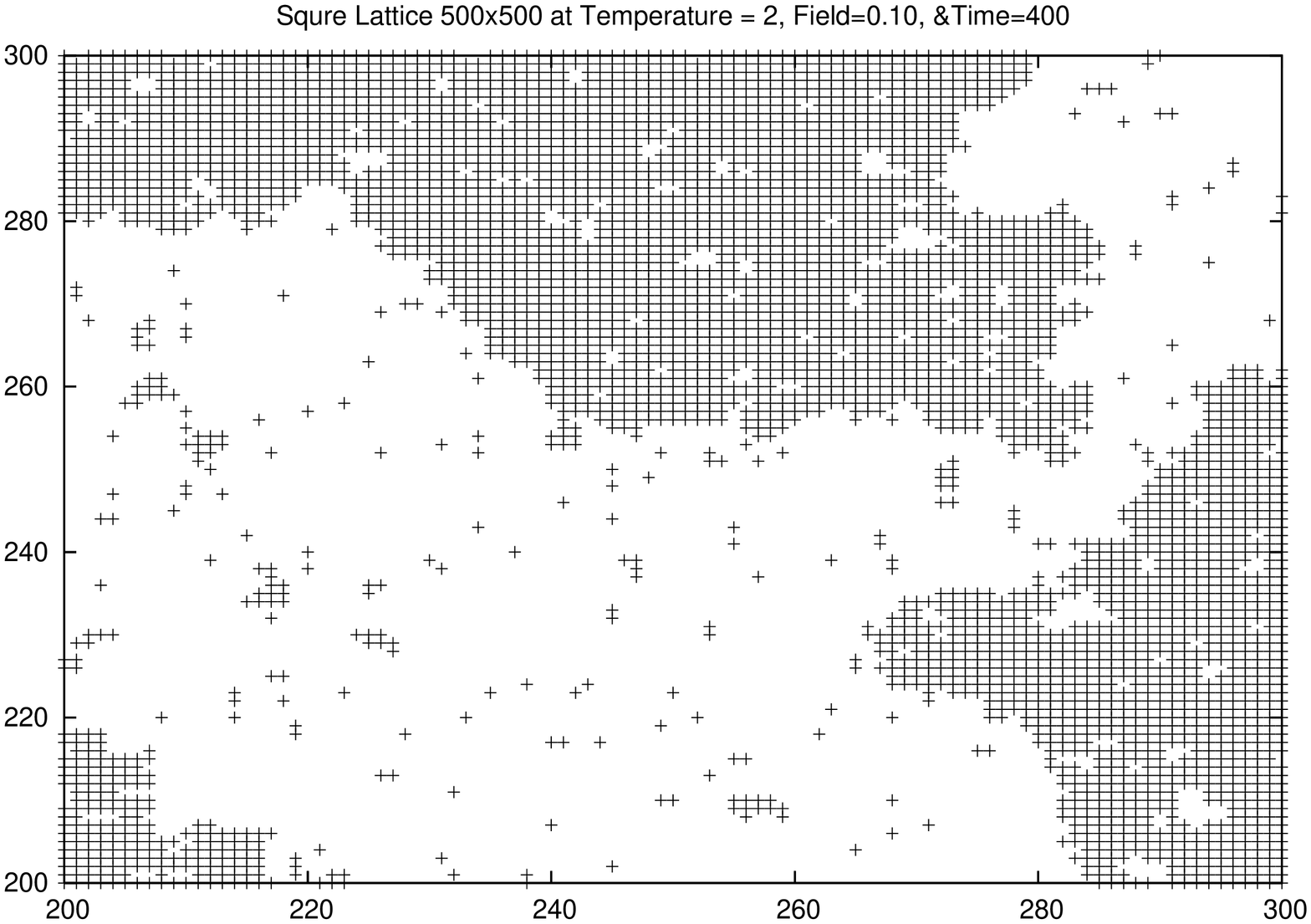}
\includegraphics [angle=-90,scale=0.3]{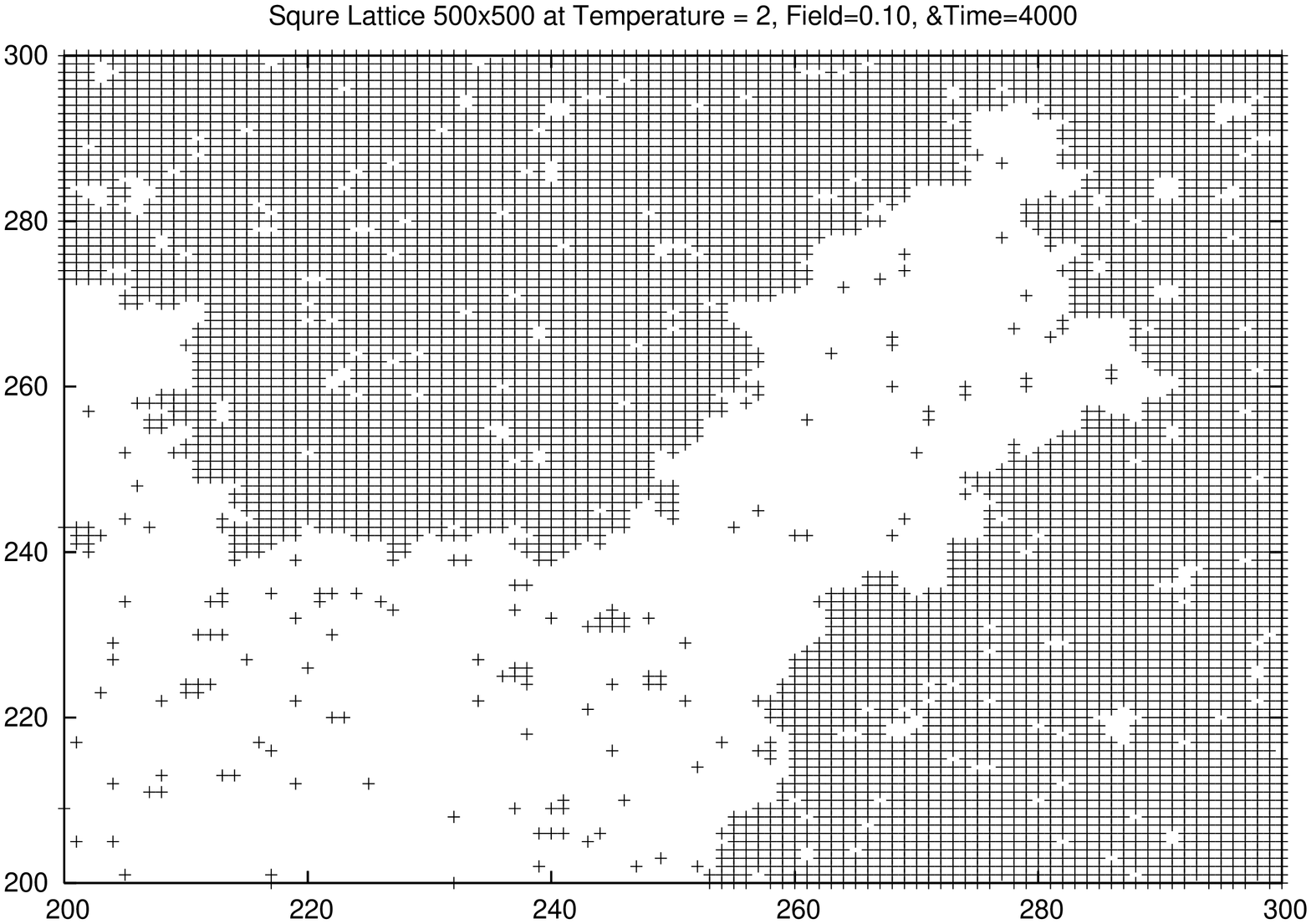}
\includegraphics [angle=-90,scale=0.3]{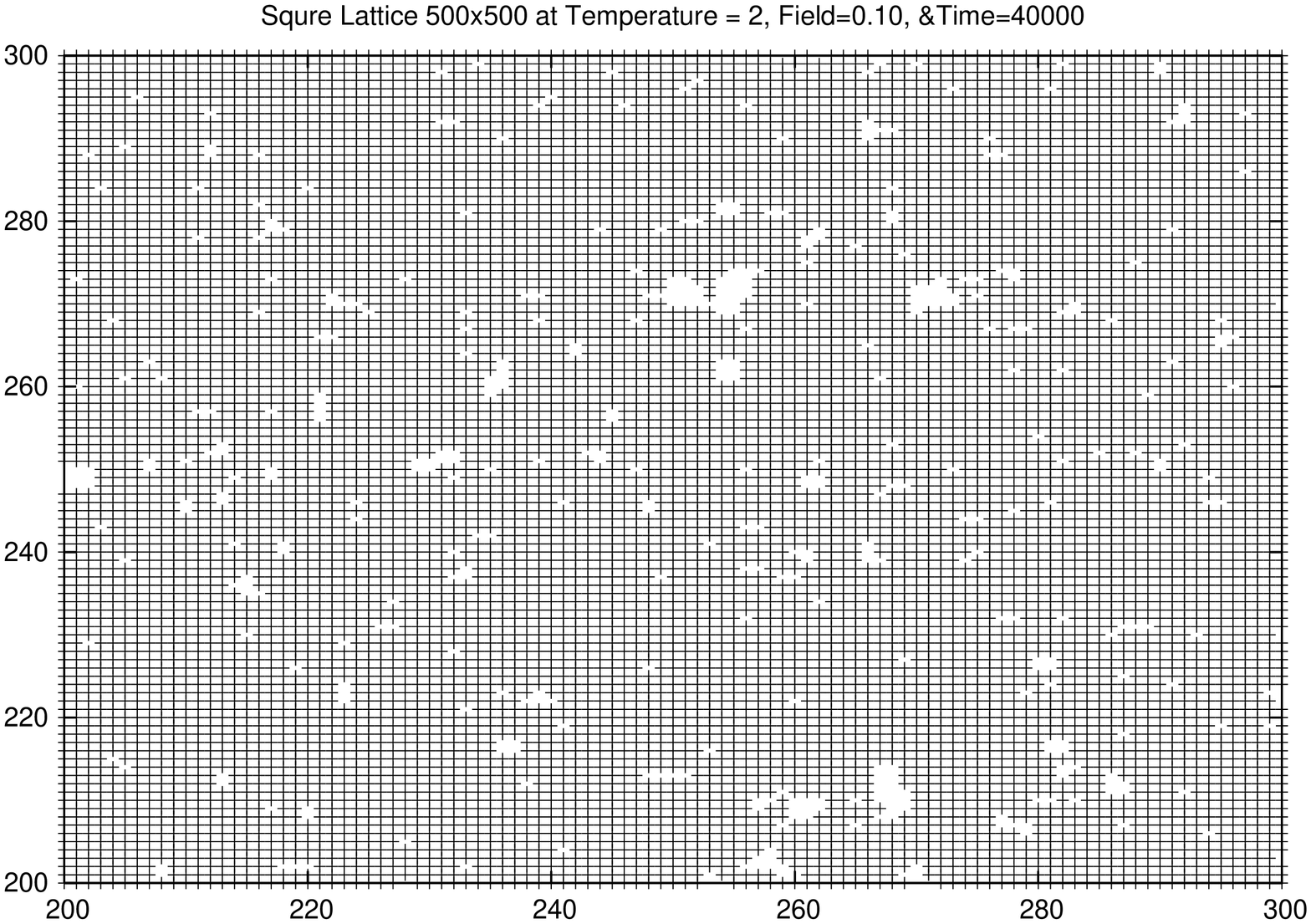}
\end{center}
\caption{Ising model simulation configurations after different times 40 (part a), 400 (part b), 4000 (part c), 40000 (part d) per site on $500 \times 500$ square lattice at $k_BT/J=2$ in a small random field of 0.1.}
\end{figure}

\bigskip
The black domains in the figures refer to rich people while the white domains refer to poor people.
We see in figure 1 from top to bottom a continuous growth of the domain size; at $t = 40,000$ we have only one domain on top and one on bottom.  In addition there are always small black clusters in the white domains and small white clusters in the black domains. They would also occur on the standard Ising model at zero field and positive temperatures: Some spins are accidentally overturned because of thermal fluctuations.

\bigskip
We then repeat the simulation of the Ising model with the same above parameters used in figure 1 but in large random fields $\pm 0.9$ and we get figure 2 (a,b,c,d).

\begin{figure}[hbt]
\begin{center}
\includegraphics [angle=-90,scale=0.3]{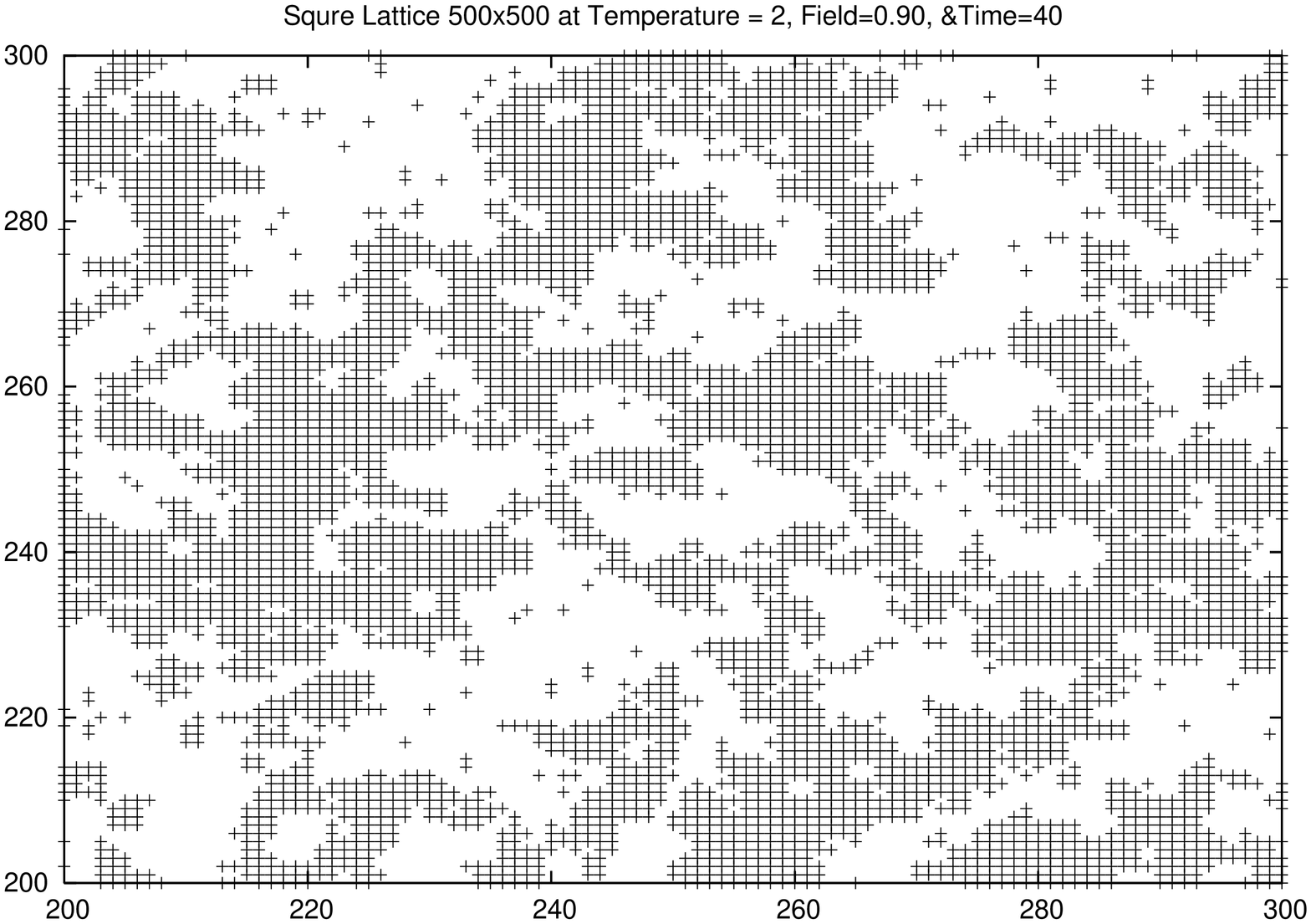}
\includegraphics [angle=-90,scale=0.3]{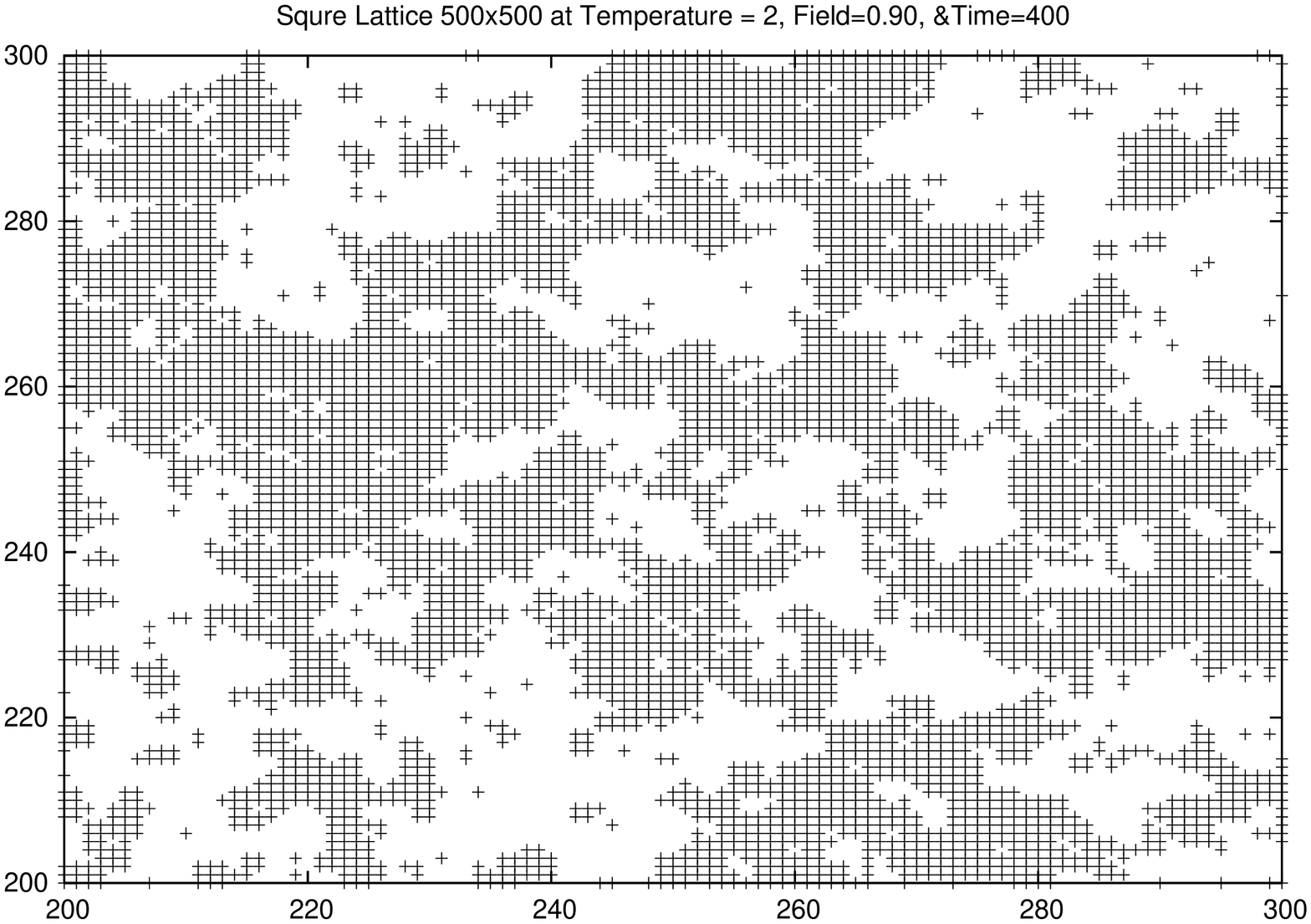}
\includegraphics [angle=-90,scale=0.3]{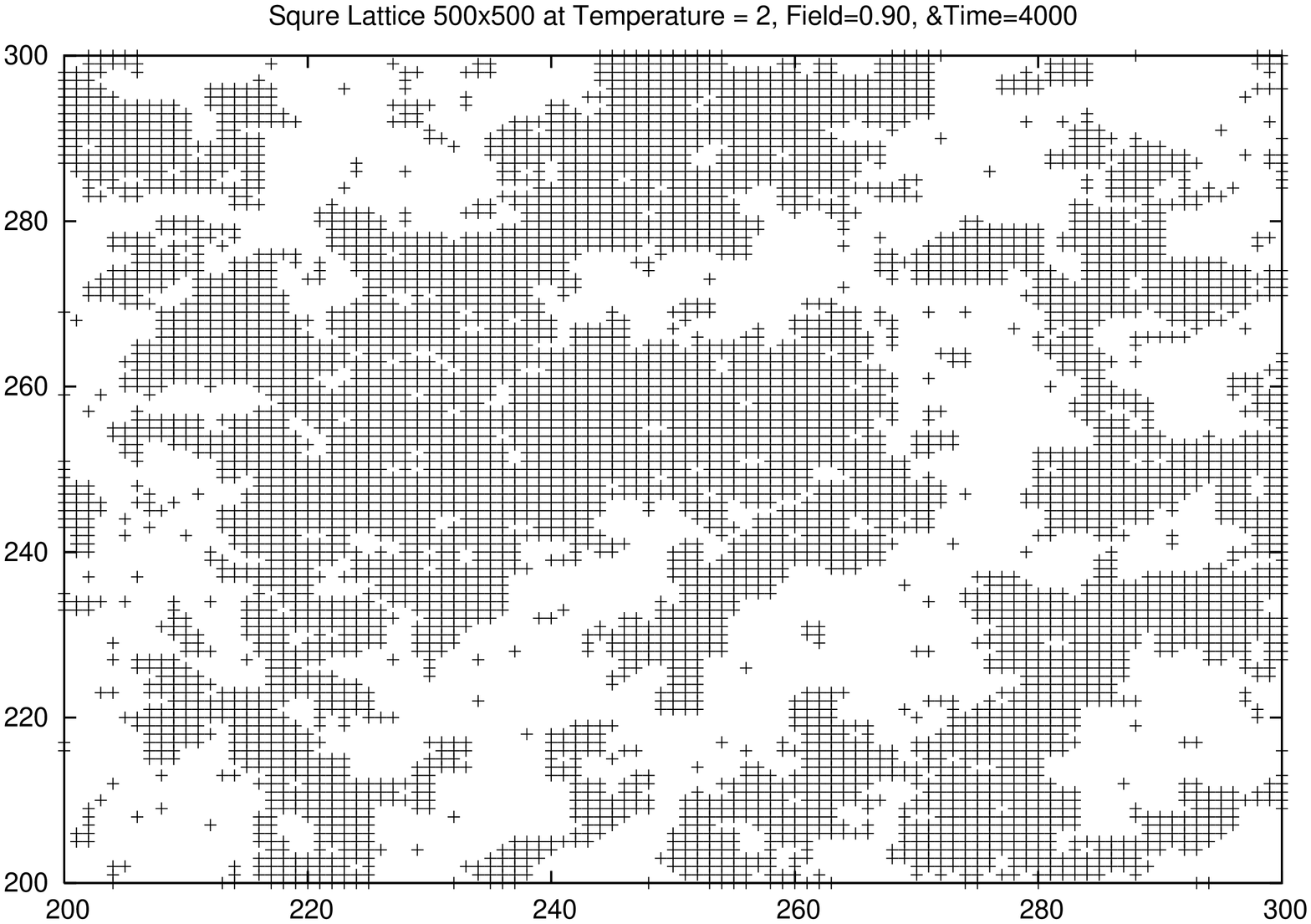}
\includegraphics [angle=-90,scale=0.3]{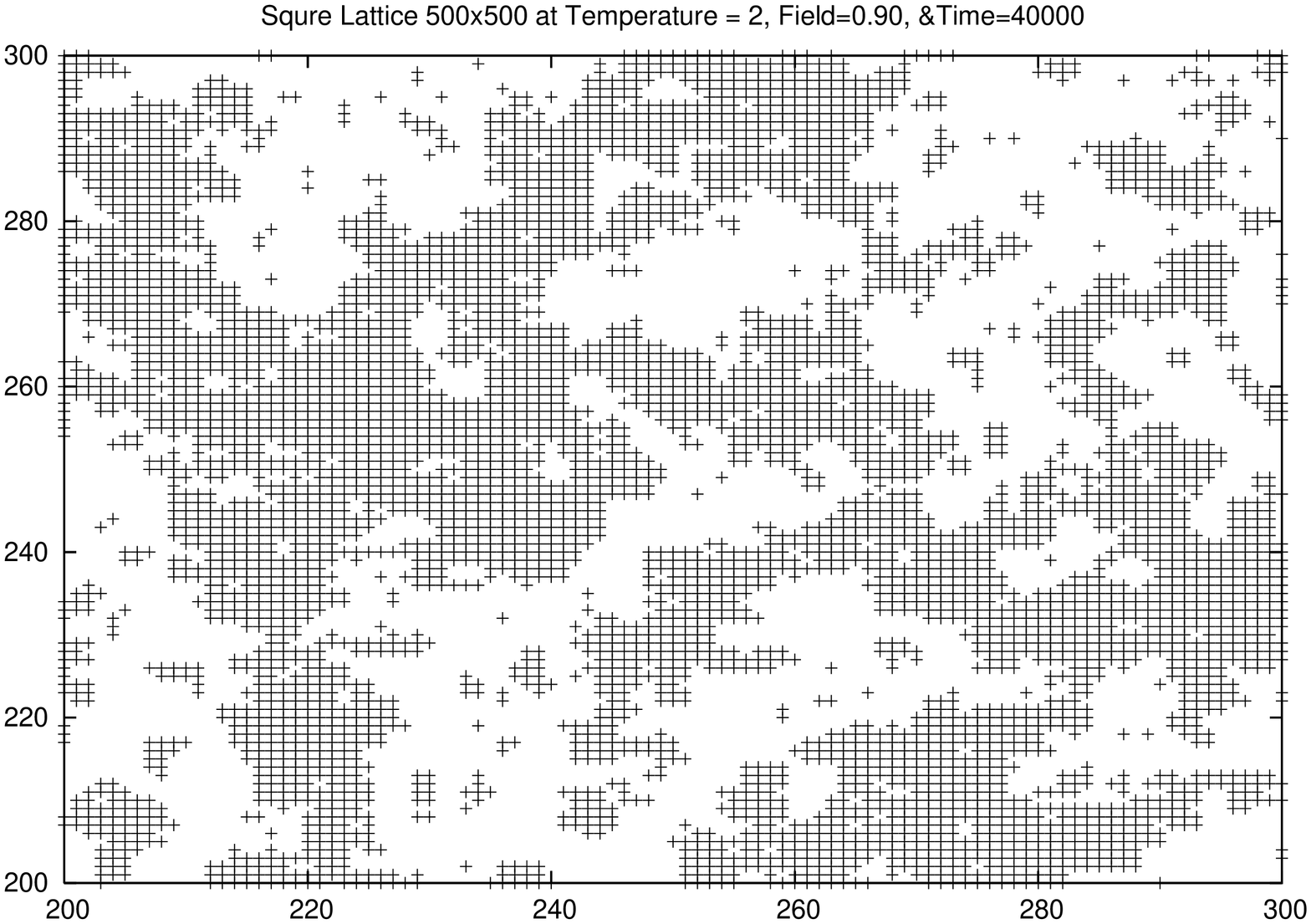}
\end{center}
\caption{Ising model simulation configurations after different times 40 (part a), 400 (part b), 4000 (part c), 40000 (part d) per site on $500 \times 500$ square lattice at $k_BT/J=2$ in a large random field of 0.9.}
\end{figure}

\bigskip
It can be noticed from figure 2 that there is strong preference of the poor to live in cheap houses and of the rich to live in expensive houses.  And since cheap and expensive houses are distributed without domains (no rich quarters) the formation of  large domains is prevented by the prices. The smaller are the price differences, the larger are the domains.

\bigskip

\centerline {\bf \large Conclusion}

We assumed that residences are either cheap or expensive, randomly distributed over the square lattice, and that two groups of people, rich and poor, make up the population. We found that for small fields after a long time the domains are larger than for large fields, in this random-field Ising model of urban segregation. Housing price differences do not prevent segregation if they are not very large.

\bigskip

\centerline {\bf \large Acknowledgment}
The Authors would like to thank D. Stauffer, Institute for theoretical Physics, Cologne University for his valuable suggestions and constructive advise.

\end{document}